# Obtención del coeficiente de fricción dinámico a partir de la aplicación de las leyes de Newton

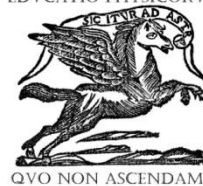

**M. López Reyes[1,2], J. Ramírez Jiménez[2]**
[1]*Departmento de investigación, Instituto Frontera. Calle Constitución No. 10,000. Colonia Centro, C.P. 22,000, Tijuana, B. C. México.*
[2]*Departamento de Física, Centro Universitario de Ciencias Exactas e Ingenierías. Universidad de Guadalajara. Blvd. Gral. Marcelino García Barragán No. 1421. Colonia Olímpica, C.P. 44430, Guadalajara, Jal. México.*

**E-mail:** mauricio.investigacion@institutofrontera.edu.mx

**Resumen**

La fuerza de fricción, coeficientes de fricción y los efectos en la dinámica de partículas, cuerpos y sistemas, son temas fundamentales en la física universitaria de los primeros ciclos y también en los cursos de física general de la educación media superior en México, sin embargo, muchos de hechos en estos temas, se dan como definición, por ejemplo; el coeficiente de fricción dinámico es menor al estático. En este trabajo, se expone una propuesta metodológica teórico-experimental, para demostrar por qué la relación entre los coeficientes es así, además se utilizaron.

**Palabras clave:** Modelado de procesos, Educación física, Enseñanza de la mecánica clásica.

**Abstract**

The force of friction, friction coefficients and the effects on the dynamics of particles, bodies and systems, are fundamental themes in university physics of the first cycles and also in general physics courses of upper secondary education in Mexico, however Many facts on these subjects are given as a definition, for example; the dynamic coefficient of friction is less than the static one. In this work, a theoretical-experimental methodological proposal is exposed, to demonstrate why the relationship between the coefficients is like this, in addition some statistical parameters were used to quantify to what extent, the experimental design improved the initial approximations.

## I. INTRODUCCIÓN

Para los profesores es claro, que los alumnos de un primer curso de física se disponen a esta asignatura con expectativas sobre la realización de actividades experimentales, que le permitan demostrar principios o teorías [1, 2]. Históricamente la física ha sido una ciencia fundamentalmente experimental, tal es así, que una gran cantidad de reconocimientos internacionales solicitan que el descubrimiento teórico sea demostrado experimentalmente. Es por ello, que la ruta didáctica seguida en los cursos de física universitaria, debe ser, aprender herramientas técnicas, como la matemática, métodos experimentales y de tratamiento de datos [3], para modelar sistemas físicos e interpretar los resultados y alcances de la teoría.

Considerando la teoría del aprendizaje significativo [4] que tiene como objetivo, estimular la curiosidad y propiciar el diálogo e intercambio de conocimiento entre los estudiantes, se ha diseñado una propuesta didáctica para modelar el comportamiento de algunas variables de un sistema masa-plano inclinado, entre ellas; el tiempo que tarda en recorrer la masa la longitud del plano inclinado, en función del ángulo o pendiente y la relación con los coeficientes de fricción estático y dinámico. Este tema forma parte de la currícula de todas las ingenierías y la licenciatura en física en los primeros semestres de la carrera, específicamente en la asignatura de mecánica, cuando se estudian las aplicaciones de las leyes de Newton. Sin embargo, lo único que los alumnos reciben como conocimiento es: "el coeficiente de fricción dinámico es menor que el estático" y las explicaciones empíricas son numerosas, no obstante las dudas siguen presentes ¿será verdad?, ¿se cumplirá para todos los materiales? ¿cuánto más grande será un coeficiente respecto del otro? ¿Cómo afectan los coeficientes la dinámica del movimiento? Aunado a estas preguntas, se establece que, el objetivo de esta propuesta didáctica, es que el estudiante cuente con un diseño teórico-experimental, para modelar e interpretar la dinámica de un sistema físico en presencia de fricción, y pueda dar una respuesta cuantitativa a las preguntas que se plantearon como motivación.

Sin duda que uno de los principales obstáculos, para el análisis e interpretación de los datos de un experimento o ecuación, que modele algún sistema físico, parte de la pobre solidez matemática [5], es por ello que, el proceso de deducción matemática se muestra con detalle. Esto con el objetivo de facilitar la interpretación física de las ecuaciones finales, así como los casos particulares y límite que de ellas se deduzcan.

Esta propuesta didáctica está recomendada para estudiantes de los primeros semestres de ingenierías y de licenciatura en física, es posible aplicarla en temas selectos de física a nivel medio superior, como fue el caso del bachillerato de Instituto Frontera, donde se aplicó a estudiantes de 5$^{\text{to}}$ semestre. Los saberes previos necesarios son, el manejo de expresiones algebraicas, razones trigonométricas, graficación de funciones y manejo de un programa de cálculo numérico, para nivel superior, se recomienda el uso de Python u Octave y en nivel medio superior, el uso de Excel genera resultados equivalentes.

## II. MATERIALES Y MÉTODO

### A. Modelo Físico. Caso estático

Como en todo curso de iniciación a la física, específicamente en la mecánica clásica, debemos partir de las leyes de movimiento de Newton. En este caso, consideremos el sistema plano inclinado-masa, que se ilustra en la figura 1 a), donde el plano inclinado se encuentra sujeto al piso horizontal sin la posibilidad de moverse, y un objeto de masa $m$, que se coloca sobre la superficie del plano inclinado, debido a las características de las superficies, existe una fuerza de fricción estática caracterizada por un coeficiente de fricción estático $\mu_s$, si consideramos el ángulo crítico en el cual el objeto sobre el plano, permanece en el límite del equilibrio traslacional, se deben de cumplir la condición vectorial (1).

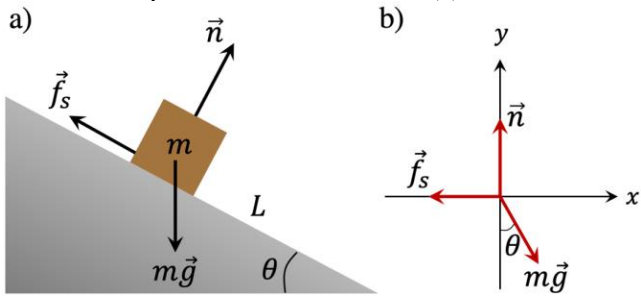

**FIGURA 1.** a) Sistema plano inclinado-masa y b) diagrama de cuerpo libre rotado $\theta$ grados.

$$\sum_i \vec{F}_i = \vec{0} \qquad (1)$$

La ecuación (1), representa la Primer Condición del Equilibrio (PCE), que es un caso particular de la segunda ley de Newton, cuando la aceleración del objeto es nula, es decir, $\ddot{\vec{r}} = \vec{0}$. [8], naturalmente, esta ecuación vectorial puede descomponerse en coordenadas cartesianas del plano (2).

$$\sum F_x = 0, \quad \sum F_y = 0. \qquad (2)$$

Teniendo en cuenta el diagrama de cuerpo libre del sistema que se presenta en la figura 1 b), y realizando la suma de fuerzas según (2) se determina, que para el caso estático, el coeficiente de fricción entre las dos superficies está dado por la expresión (3).

$$\mu_s = \tan \theta. \qquad (3)$$

De la definición de fuerza de fricción estática, bien conocida por los estudiantes, se sabe que, esta depende del coeficiente de fricción estático y de la fuerza normal $\vec{n}$, implícitamente de la masa del objeto y del ángulo de la superficie, algebraicamente se define como en (4).

$$\vec{f}_s = \mu_s \vec{n}. \qquad (4)$$

La expresión para la fuerza de fricción dinámica es equivalente a (4), con la diferencia numérica, que el coeficiente de fricción dinámico es menor al estático [6] y por lo tanto, la magnitud de la fuerza de fricción dinámica es menor a la estática.

### B. Caso dinámico (con fricción estática)

Para el caso dinámico, es preciso determinar el ángulo crítico en el cual, idealmente el objeto permanecerá en equilibrio traslacional, es decir, en reposo relativo al plano inclinado, o bien, en movimiento con velocidad constante. Al incrementar la pendiente del plano, la fuerza de fricción estática será menor que la componente horizontal del peso por lo que el objeto adquiere aceleración, ver figura 2.

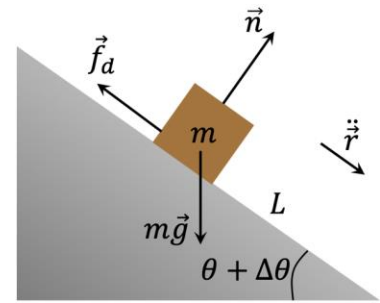

**FIGURA 2.** Sistema plano inclinado-masa para el caso dinámico. El diagrama de cuerpo libre tiene la misma forma.

La segunda ley de Newton para el caso dinámico se escribe como,

$$\sum_i \vec{F}_i = m\ddot{\vec{r}}. \qquad (5)$$

Al descomponer esta ecuación es sus componentes cartesianas, obtenemos las expresiones para la suma de fuerzas en el eje horizontal y vertical respectivamente en (6)

$$\left. \begin{array}{rcl} mg \sin \theta - f_s & = & ma_x \\ n - mg \cos \theta & = & 0 \end{array} \right\}. \qquad (6)$$

Resolviendo el sistema (6) para la aceleración, tomando en cuenta (4), se llega a la solución analítica:

$$a_x = g\left(\sin\theta - \mu_s \cos\theta\right). \qquad (7)$$

Se observa que la aceleración depende del ángulo de inclinación de la rampa y del coeficiente de fricción, en este caso, se utilizó el coeficiente estático encontrado por (3), posteriormente se utilizará para determinar el coeficiente de fricción dinámico cuyo modelo es equivalente a (7). Es por ello que se puede considerar a la aceleración como una función que depende de 2 variables independientes, $a_x(\theta, \mu_s)$. En la figura 3, se observa el comportamiento de la aceleración en función del ángulo de inclinación y del coeficiente de fricción, el dominio de esta función es:
$Dom\ a_x(\theta, \mu_s) = \left\{(\theta, \mu_s) \in \mathbb{R}^2 : 0 \leq \theta \leq \frac{\pi}{2},\ 0 \leq \mu_s < 1\right\}$

Retomando la expresión (7) y la gráfica de la figura 3, tomemos los siguientes casos límites:

i.   Si $\mu_s \to 0$, entonces,  $\qquad a_x \to g\sin\theta$

ii.  Si $\mu_s \to 0$, $\theta \to 90°$, entonces, $\qquad a_x \to g$

iii. Si $g \to 0\ m/s^2$, entonces, $\qquad a_x \to 0\ m/s^2$

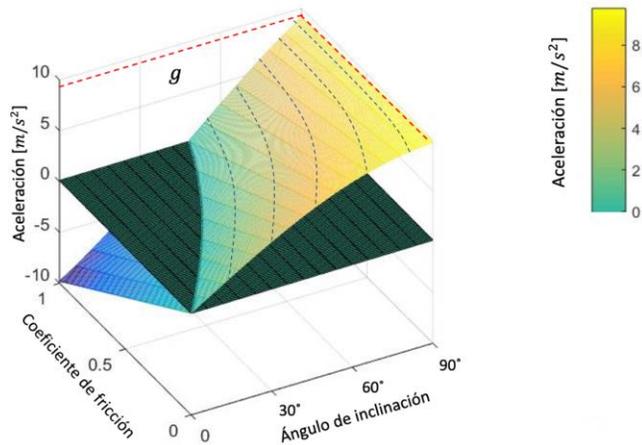

**FIGURA 3.** Gráfica de la función (7) donde se observa el comportamiento de la aceleración en función del ángulo y el coeficiente de fricción. NOTA: El eje del ángulo debe estar en grados y sólo se debe ver la parte superior al plano z=0.

Obsérvese como los primeros dos casos, corresponden a una realidad física en presencia de aceleración gravitacional, donde se elimina la fuerza de fricción y se inclina verticalmente la rampa, simplificando la situación a caída libre sin rozamiento. En el caso iii, la simplificación es tal que, no hay aceleración gravitacional, por lo que la aceleración del objeto sobre el plano inclinado es nula, en otras palabras, no importa el ángulo de inclinación ni el coeficiente de fricción.

## C. Cinemática del movimiento (con fricción estática)

De la definición de aceleración, $\ddot{\vec{r}} = \vec{a}$, para el caso unidimensional, tenemos $\ddot{x} = a_x$, cuya solución particular con las condiciones iniciales; $\dot{x}(0) = v_i$, $x(0) = x_i$, es la expresión siguiente

$$x_f = \frac{1}{2}a_x t^2 + v_i t + x_i. \qquad (8)$$

Para el caso de nuestro modelo, el objeto parte del reposo, por lo que $v_i = 0\ m/s$ y la cantidad $x_f - x_i$ es la longitud de la rampa $L$. De esta manera la ecuación (8) se reduce a:

$$\frac{1}{2}a_x t^2 = L. \qquad (9)$$

Sustituyendo (7) en (9) y resolviendo para el tiempo, obtenemos:

$$t = \sqrt{\frac{2L}{g(\sin\theta - \mu_s \cos\theta)}}\ . \qquad (10)$$

En la figura 4, se observa el comportamiento del tiempo que tarda el objeto en recorrer el plano inclinado en función del coeficiente de fricción, observar que el coeficiente de fricción estático se obtiene de (3) y aproxima en primera instancia al tiempo teórico (10), para cuestiones de graficación, se considera $g = 9.8\ m/s^2$, $L = 0.8\ m$ y diversos ángulos en el intervalo (0,90°). En el desarrollo experimental, se realizan mediciones experimentales para ver que tan buena es la aproximación del tiempo de caída usando el coeficiente de fricción estático.

Analizando los casos límite de (10) podemos realizar las siguientes simplificaciones e interpretaciones físicas:

i.   Si $L \to \infty$, entonces, $\qquad t \to \infty$,

ii.  Si $L \to 0$, entonces, $\qquad t \to 0$,

iii. Si $\mu_s \to 0$, entonces, $\qquad t \to \sqrt{\frac{2L}{g\cos\theta}}$,

iv.  Si $\mu_s \to 0$, $\theta \to 90°$, entonces, $\qquad t \to \sqrt{\frac{2L}{g}}$.

La primera y segunda situación no requiere explicación abundante, es obvio que el tiempo de caída del objeto depende de la longitud del plano. La condición iii, nos reduce la cinemática del movimiento a caída libre con el valor efectivo de la aceleración gravitacional, $a_x = g\cos\theta$, dado por la geometría del plano. El caso límite iv, simplifica nuestro movimiento a caída libre, fácilmente se observa que la expresión para el tiempo se reduce a (9). Este tipo de análisis de casos límites son interesantes y muy útiles durante el proceso de enseñanza-aprendizaje, ya que fomentan el pensamiento crítico en los estudiantes, interpretando la situación física desde los casos generales, hasta particularidades, cuya comprobación experimental es más intuitiva [7].

**D. Obtención del coeficiente de fricción dinámico**

Para el caso dinámico, el modelo físico es idéntico que para el caso estático, salvo la fuerza y coeficiente de fricción. En la ecuación (10) se estimó el tiempo que tarda en recorrer la rampa el objeto, considerando el coeficiente de fricción estático. Debería ser claro para el profesor que el tiempo teórico debe ser mayor que el tiempo experimental $t_e$, ya que la fuerza de fricción estática utilizada es mayor que la dinámica (no conocida), $f_s > f_d$, por lo tanto, el valor de la aceleración del objeto es menor y en consecuencia el tiempo teórico es más grande. Esta premisa, es una hipótesis que el estudiante debe plantear, en donde refleja la comprensión de los conocimientos teóricos previos, sin embargo, aún no se responden totalmente las preguntas planteadas inicialmente. Utilizando la expresión (10), cambiando el término del coeficiente de fricción estático por el dinámico y despejando este último, se obtiene la ecuación:

$$\mu_d = \tan\theta - \frac{2L}{gt_e^2 \cos\theta}. \quad (11)$$

Observe el resultado del coeficiente de fricción dinámico, depende en gran medida de la capacidad que se tenga para medir con precisión el ángulo de inclinación y el tiempo. Para llevar a cabo la recolección de estos datos experimentales, se necesita de una cámara de video que permita reproducir el video con menor rapidez. En la figura 4 se observa el diseño experimental.

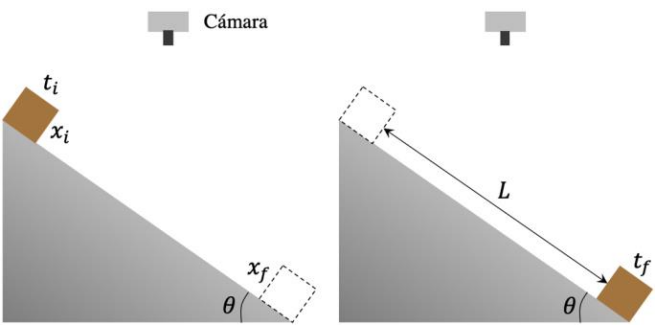

**FIGURA 4.** Diseño experimental que consta de un plano inclinado con la facilidad para cambiar los ángulos y marcado con la posición inicial y final, un bloque y cámara de video.

El desarrollo experimental es el siguiente:
1. Determinar el coeficiente de fricción estático con base en (3).
2. A partir del ángulo crítico obtenido en 1, iniciar la toma de datos para el tiempo experimental, incrementando el ángulo de inclinación 5° para cada medición.
3. Repetir 3 veces las mediciones del punto 2, y calcular el tiempo experimental medio, así como la desviación estándar.
4. Comparar los resultados experimentales, con la aproximación (10), utilizando el coeficiente de fricción estático.
5. A partir de los tiempos experimentales medios y con (11), encontrar el valor del coeficiente de fricción dinámico experimental así como su desviación estándar.
6. Comparar los resultados obtenidos con la aproximación del coeficiente de fricción estático contra el dinámico.

## IV. RESULTADOS EXPERIMENTALES

Como ejemplo de la aplicación de la propuesta didáctica, consideremos un diseño como el de la figura 4, donde la longitud $L = 80.0 \pm 0.05\ cm$. Se determinó que el ángulo crítico fue de 16°, por lo que con (3), el coeficiente de fricción estático entre las superficies que se utilizaron es de 0.29 (bloque de plásticos sobre rampa de acrílico).

La toma de datos del tiempo experimental se presenta en la tabla I, donde se observan las tres mediciones obtenidas, el tiempo experimental medio y la desviación estándar de cada una de ellas.

**TABLA I.** Mediciones del tiempo experimental, tiempo experimental medio (en negritas) y desviación estándar.

| Ángulo [°] $\pm 0.05°$ | Tiempos registrados [s] $\pm 0.005s$ | | | | $\pm \sigma, [s]$ |
|---|---|---|---|---|---|
| | $t_{e1}$ | $t_{e2}$ | $t_{e3}$ | $t_e$ | |
| 20 | 1.34 | 1.30 | 1.27 | **1.30** | 0.035 |
| 25 | 0.96 | 0.90 | 0.94 | **0.93** | 0.031 |
| 30 | 0.80 | 0.81 | 0.82 | **0.77** | 0.010 |
| 35 | 0.68 | 0.63 | 0.65 | **0.65** | 0.025 |
| 40 | 0.60 | 0.58 | 0.62 | **0.60** | 0.020 |
| 45 | 0.55 | 0.53 | 0.54 | **0.54** | 0.010 |
| 50 | 0.51 | 0.51 | 0.53 | **0.52** | 0.006 |
| 55 | 0.49 | 0.48 | 0.46 | **0.48** | 0.015 |
| 60 | 0.47 | 0.47 | 0.46 | **0.47** | 0.006 |
| 65 | 0.44 | 0.46 | 0.44 | **0.45** | 0.006 |
| 70 | 0.45 | 0.44 | 0.43 | **0.44** | 0.006 |
| 75 | 0.42 | 0.41 | 0.42 | **0.42** | 0.006 |
| 80 | 0.41 | 0.42 | 0.42 | **0.42** | 0.006 |
| 85 | 0.41 | 0.41 | 0.40 | **0.41** | 0.006 |
| 90 | 0.40 | 0.41 | 0.40 | **0.40** | 0.006 |

En la figura 5 se presenta la gráfica de los tiempos experimentales en función del ángulo de inclinación.

Es interesante comparar los tiempos experimentales medios, con los teóricos de la ecuación (10) en la aproximación del coeficiente de fricción estático.

Con base en los resultados experimentales de la tabla I y la ecuación (11), se estimó el coeficiente de fricción dinámico para cada uno de los promedios de los tiempos $t_e$, resultados que se presentan en la tabla II.

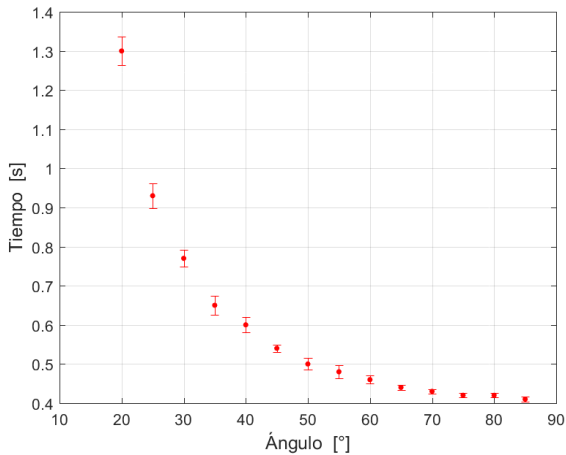

**FIGURA 5.** Datos experimentales del tiempo en recorrer la rampa en función del ángulo de inclinación. Se observa la desviación estándar con los bigotes alrededor del punto medio.

Se determinó con los datos de la tabla II que, el valor medio del coeficiente de fricción dinámica $\bar{\mu}_d$, es de 0.25, con una desviación estándar de $\pm 0.042$.

**TABLA II.** Estimación del coeficiente de fricción dinámica, a partir de los datos experimentales de la tabla I.

| Ángulo [°] $\pm 0.05°$ | $t_e$ [s] $\pm 0.005\,s$ | $\mu_d$ |
|---|---|---|
| 20 | 1.30 | 0.27 |
| 25 | 0.93 | 0.26 |
| 30 | 0.77 | 0.26 |
| 35 | 0.65 | 0.24 |
| 40 | 0.60 | 0.25 |
| 45 | 0.54 | 0.21 |
| 50 | 0.52 | 0.25 |
| 55 | 0.48 | 0.19 |
| 60 | 0.47 | 0.25 |
| 65 | 0.45 | 0.24 |
| 70 | 0.44 | 0.28 |
| 75 | 0.42 | 0.16 |
| 80 | 0.42 | 0.33 |
| 85 | 0.41 | 0.29 |
| 90 | 0.40 | N/D |

En la figura 6, se observa la comparativa entre los tiempos experimentales reales y el modelo, utilizando los coeficientes de fricción estático y dinámico correspondientes a las líneas continuas y a trozos, respectivamente. Se vuelve evidente que el modelo matemático de (10) describe correctamente el comportamiento del tiempo en función del ángulo de inclinación, tomando en cuenta el coeficiente de fricción, además que para el caso del coeficiente de fricción dinámico, el ajuste es mejor que para el caso dinámico, utilizando el coeficiente de correlación $R^2$ para ambos ajustes, se determinó que para el caso estático (línea continua), $R_s^2 = 0.990$, y para el caso dinámico $R_d^2 = 0.999$, lo que permite demostrar que el diseño experimental, mejora la aproximación de la dinámica del sistema.

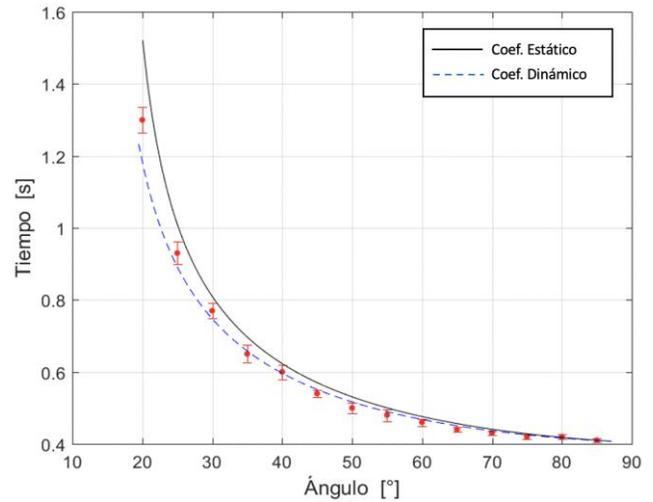

**FIGURA 6.** Tiempos experimentales en función del ángulo de inclinación de la rampa y curvas teóricas del tiempo en función del ángulo, para la curva continua se tomó en cuenta el coeficiente de fricción estático y para la línea a trozos el coeficiente de fricción dinámico.

### A. Evaluación y análisis de los resultados

Con base en la tabla II, es notorio que no todos los datos convergen a la media, tal es el caso registrado en los ángulos de 55°, 75° y 80° en donde claramente se observa que los valores del $\mu_d$, se alejan de la media, posibles explicaciones para esta discrepancia en los datos son las siguientes; errores en la toma de los tiempos experimentales, desgaste o suciedad de las superficies tanto de la rampa como del bloque y pensando en el desarrollo de este experimento en el aula, pueden influir otros errores sistemáticos y aleatorios. Lo anterior no necesariamente constituye un problema o el fracaso del experimento ya que estos resultados pueden ser aprovechados por el conducto docente para contrastar la teoría con la realidad y proponer mejorar el diseño del experimento, recolección y procesamiento de datos.

Sin duda que el objetivo principal, que es demostrar que el coeficiente de fricción dinámico es menor que el estático se cumplió con relativo éxito, cabe mencionar que con base en los casos límite de (10) se observa que para el ángulo de 90°, la situación física se reduce a caída libre, razón por la cual no tiene sentido determinar el coeficiente de fricción dinámico en ese ángulo (no hay contacto entre las superficies) la fuerza normal es cero. Situaciones como esta, pueden constituir una valiosa oportunidad para el docente de evaluar según la taxonomía de Bloom, el nivel de la complejidad del aprendizaje.

Los resultados de este experimento, son un área de oportunidad para el docente de volver significativo el aprendizaje [4], ya que en el desarrollo instruccional, tanto de la teoría, como de la experimentación, se privilegia el razonamiento cuantitativo-deductivo, por ejemplo; en los

*M. López Reyes & J. Ramírez Jiménez*

casos límite de (10), se plantean situaciones particulares e hipotéticas que ocurrirían en el experimento a partir de una ecuación general, el rol del docente de física, será inducir al alumno desde la abstracción de las ecuaciones y principios fundamentales, hasta la modelación o aplicación de situaciones particulares.

Particularmente hablando de los coeficientes de fricción estático y dinámico, la propuesta didáctica expuesta, permite que el alumno concluya con bases experimentales que, al menos el coeficiente de fricción dinámico de los materiales involucrados es menor que el estático y sus consecuencias se pueden modelar y contrastar con la teoría.

## VI. CONCLUSIONES Y RECOMENDACIONES

Si bien, la aproximación de la dinámica del movimiento es bastante buena utilizando el coeficiente de fricción estático (para este caso particular), con la refinación de este valor utilizando el arreglo experimental propuesto, los resultados mejoran a tal grado que, se observa una correlación casi perfecta entre los datos experimentales y el modelo.

Este experimento permite además de profundizar en los temas de la dinámica de Newton, abordar el problema de las cifras significativas y la teoría de la incertidumbre, por lo que se podría considerar como un ejercicio integrador, tanto de la teoría física, como de los métodos y prácticas experimentales.

En la práctica, es probable que no se cuente con instrumentos de alta precisión, especialmente si se trata de un laboratorio de docencia o incluso, un salón de clases es por ello, por lo que este diseño experimental tiene la versatilidad de poder realizarse únicamente con la cámara y cronómetro de un teléfono celular, sin sacrificar perdida de exactitud considerable, comparada con la propia incertidumbre de las cifras significativas.

La implementación de experimentos y modelación constituye una herramienta complementaria muy valiosa para la física, especialmente para la mecánica clásica ya que dota de sentido y vuelve significativos los aprendizajes.

Se recomienda replicar este diseño experimental con otros materiales, e identificar aquellas situaciones en que su funcionamiento es mejor, por ejemplo, en cierto intervalo de ángulo de inclinación.

**AGRADECIMIENTOS**

Los autores agradecemos al **Departamento de Investigación Educativa de Instituto Frontera** por el apoyo recibido durante el periodo en que duró la investigación, así como a los alumnos de dicha institución, mismos que realizaron la validación experimental con diversos materiales. Además, también agradecemos a la Universidad Regional de los Altos **URA**, por facilitarnos alumnado y materiales de laboratorio.
**REFERENCIAS**

[1] Halloun, I., *Views about science and physics achievement: The VASS story*, In E.F. Redish and J. Rigden (Eds.), AIP Conference Proceedings **399**, The changing role of physics departments in modern universities (pp. 605-614), American Institute of Physics (1997).

[2] Perkins, K. K., Adams, W. K., Pollock, S. J., Finkelstein, N. D., and Wieman, C.E., *Correlating student beliefs with student learning using the Colorado Learning Attitudes about Science Survey*, In J. Marx, P. Heron, and S. Franklin (Eds.), AIP Conference Proceedings **790**, 2004 Physics Education Research Conference (pp. 61-64). Melville, NY: American Institute of Physics (2005).
https://aip.scitation.org/doi/abs/10.1063/1.2084701

[3] Wheeler, G. F., *Three faces of inquiry*, In J. Minstrell & E. H. van Zee (Eds.), Inquiring into inquiry learning and teaching in science, (pp. 14–19). Washington, DC: American Association for the Advancement of Science (2000).

[4] Moreira, M. A., *Aprendizagem Significativa Crítica, 2ª* Edição, (Editora Plátano, Lisboá, 2010).

[5] Monk, M., *Mathematics in physics education: a case of more haste less speed*, Phys. Educ. **29**, 209-211 (1994).

[6] Sears, F. W., Zemansky, M. W., Young, H. D., Freedman, R. A*., Física Universitaria* Vol. 1 (Editorial Pearson, México, 2013), pp. 120-123.

[7] Ilzarbe, L., Tanco, M., Viles, E., Álvarez, M., *El diseño de experimentos como herramienta para la mejora de los procesos. Aplicación de la metodología al caso de una catapulta,* (Tecnura, Bogotá, 2007), pp. 128-129.
https://www.redalyc.org/pdf/2570/257021012011.pdf

[8] Sears, F. W., Zemansky, M. W., Young, H. D., Freedman, R. A*., Física Universitaria* Vol. 1 (Editorial Pearson, México, 2013), pp. 104-119.